\documentclass[aps,prc,twocolumn,superscriptaddress,showpacs,nofootinbib,floatfix]{revtex4}
\usepackage{graphicx}

\newcommand{\affCENBG}{\affiliation{Centre d'\'Etudes Nucl\'eaires de Bordeaux Gradignan - 
Universit\'e Bordeaux 1 - UMR~5797~CNRS/IN2P3, Chemin du Solarium, BP 120, 33175 Gradignan, France}}
\newcommand{\affAlger}{\affiliation{Facult\'e de Physique, USTHB, B.P.32, El Alia, 16111 Bab Ezzouar, Alger, Algeria }}
\newcommand{\affMadrid}{\affiliation{Instituto de Estructura de la Materia, CSIC, Serrano 113bis, E-28006-Madrid, Spain}}
\newcommand{\affGANIL}{\affiliation{Grand Acc\'el\'erateur National d'Ions Lourds, B.P. 55027, F-14076 Caen Cedex 05, France}}
\newcommand{\affAarhus}{\affiliation{Department of Physics and Astronomy, University of Aarhus, Ny Munkegade 1520, DK-8000 Aarhus C, Denmark}}
\newcommand{\affGSI}{\affiliation{Gesellschaft f\"ur Schwerionenforschung mbH, Planckstrasse 1, D-64291 Darmstadt, Germany}}

\begin{document}
\renewcommand{\textfraction}{.01}
\headsep 1.cm

\title{Detailed $\beta$-decay study of $^{33}$Ar}

\author{N. Adimi} \affCENBG \affAlger
\author{R. Dom\'{\i}nguez-Reyes} \affMadrid
\author{M. Alcorta} \affMadrid
\author{A. Bey} \affCENBG
\author{B. Blank} \affCENBG
\author{M.J.G. Borge} \affMadrid
\author{F. de Oliveira Santos} \affGANIL
\author{C. Dossat} \affCENBG
\author{H.O.U. Fynbo} \affAarhus
\author{J. Giovinazzo} \affCENBG
\author{H.H. Knudsen} \affAarhus
\author{M. Madurga} \affMadrid
\author{I. Matea} \affCENBG \altaffiliation{Permanent address: Institut de physique nucl\'eaire d'Orsay, 5 rue Georges Cl\'emenceau, F-91406 Orsay, France}
\author{A. Perea} \affMadrid
\author{K. S{\"u}mmerer} \affGSI
\author{O. Tengblad} \affMadrid
\author{J.C. Thomas} \affGANIL

\begin{abstract}
The proton-rich nucleus $^{33}$Ar has been studied by detailed proton and $\gamma$-ray spectroscopy at the
low-energy facility of SPIRAL at GANIL. Proton and $\gamma$-ray singles and coincidence measurements allowed to establish a quasi
complete decay scheme of this nucleus. By comparing the proton intensity to different daughter states, tentative 
spin assignments have been made for some of the states of $^{33}$Cl. The Gamow-Teller strength distribution is 
deduced and compared to shell-model calculations and a quenching factor is determined. States close to the 
isobaric analogue state are searched for with respect to isospin mixing.
\end{abstract}

\pacs{ {21.10.-k} {Properties of nuclei}, {23.50.+z} {Decay by proton emission}, {29.30.Ep} {Charged-particle spectroscopy}}

\maketitle

\section{Introduction}

An important part of information about the structure of the atomic nucleus is obtained by nuclear $\beta$ decay.
Studies close to the valley of nuclear stability enable to precisely test our understanding of this structure, 
whereas experiments and theoretical work with nuclei far away from the stability line enables investigations 
of the evolution of nuclear structure as a function of isospin. However, further away from stability and in particular
on the proton-rich side of the valley of stability, nuclei no longer
decay by simple $\beta$ decay, but by $\beta$-delayed particle emission~\cite{blank08review}. Therefore,
spectroscopic studies become more difficult, as $\gamma$-ray spectroscopy has to be combined with particle detection, 
i.e. proton and $\alpha$-particle detection.

Gamma detection is commonly achieved with high-efficiency germanium detectors which allow to resolve different $\gamma$ lines
with good precision. Charged particles are detected with silicon detectors which can reach modest energy resolutions of about
10-20 keV. A combination of charged-particle detection and $\gamma$-ray measurements may permit to establish even
complicated $\beta$-decay schemes and compare them to theoretical predictions from microscopic models as the nuclear
shell model.

A long-standing problem is the observation of "quenching" of the Gamow-Teller (GT) strength, when experimental results, either
from nuclear $\beta$ decay or from charge-exchange reactions, are compared with theoretical results.  Having
a slight mass dependence this quenching reaches a value of about 0.6 for the transition strength. The quenching is seen from a
comparison of the experimental and the theoretically calculated transition strength B(GT)~\cite{brown85}.

Although this quenching is known since many years and has been observed for basically all nuclei where these kinds of 
studies has been performed, two possible explanations are still discussed in the literature (see~\cite{arima99}): i) problems
in the shell-model calculations which do not take into account enough orbitals to describe the transition strength
correctly. In particular, intruder orbitals giving rise to higher-energy excitations are usually not included to
keep the model calculations feasible. ii) Sub-nuclear excitations, especially of the $\Delta$ resonance, may shift
transition strength to excitation energies as high as 100~MeV and are therefore rather difficult to observe experimentally.
A paper of Arima~\cite{arima99} retracing to some extent the history of this quenching shows that the GT strength quenching is
to a large extent due to two-particle two-hole excitations not included in most calculations.

Another question of debate is the isospin mixing or isospin purity of nuclear states. In the $\beta$ decay of proton-rich
nuclei, the Fermi transition populates the isobaric analogue state (IAS) of the decaying ground state. For $\beta$-emitting
nuclei with isospin projection $T_z < $ 0, this IAS is within the Q$_{EC}$ window and decays either by $\gamma$ emission or, for
sufficiently proton-rich nuclei, by proton emission. However, as this IAS has an isospin quantum number equal to the one
of the parent nucleus, proton emission from this state to states in the $\beta$-delayed proton ($\beta$p) emission daughter
nucleus is forbidden by isospin conservation and thus only permitted by a small isospin impurity of either
the IAS or of the final state after proton emission. As the IAS usually lies in a region with a much higher density of
nuclear states than the daughter state, often a nuclear ground state, the IAS has generally a higher degree of isospin impurity than
its daughter state. If this is the case, the isospin mixing has its origin mainly in a mixing of the IAS with nearby 
lying states having a one-unit lower isospin quantum number, but the same spin quantum number. According to perturbation
theory, the mixing gets larger with decreasing energy difference between the two states. Of course, mixing is possible not
only with one state, but with any state, however, with rapidly decreasing amplitude as the energy difference increases.

The decay of $^{33}$Ar, the subject of the present work, was studied several times in the past 
(see~\cite{reeder64,hardy64,poskanzer66,hardy71,borge87,schardt93,honkanen96,garcia00}). 
Our experimental data are compared to the most recent work from these authors and experimental averages 
are confronted to theoretical shell-model calculations using different effective interactions optimized for the 2s1d 
nuclear shell-model space. In particular, we will determine the Gamow-Teller strength distribution and investigate
the possibility to identify the state(s) responsible for isospin mixing in $^{33}$Cl. The isospin mixing 
of the IAS was suggested to be as much as 20\% for $^{33}$Cl~\cite{hardy71}, a huge value compared to "today" standards.

Hardy et al.~\cite{hardy71} and Borge et al.~\cite{borge87} studied the decay of $^{33}$Ar in detail in order to extract the GT strength 
distribution. Schardt et al.~\cite{schardt93} and Garcia et al.~\cite{garcia00} also determined limits of possible scalar contributions
to $\beta$ decay and the Fermi-to-Gamow-Teller ratio of the decay of $^{33}$Ar. We will in particular use this ratio in our
analysis and comparison to theory. 

In the present work, the use of high-resolution silicon surface-barrier detectors and of a high-efficiency $\gamma$-detection array 
permitted to determine the proton branches to the ground as well as to the first and second excited states in the proton daughter.
In particular, the GT distribution in the full Q$_{EC}$ window was established and a detailed comparison to shell-model calculations
was performed.

\section{Experimental details}

The isotopes of interest were produced by projectile fragmentation of a $^{36}$Ar$^{18+}$ 
primary beam at 95~MeV/nucleon with intensities between 4~$\mu$A and 8~$\mu$A provided by the coupled 
cyclotrons of GANIL. This beam was fragmented in the SPIRAL carbon target. The SPIRAL ECR source then produced
a low-energy beam of $^{33}$Ar$^{3+}$, which was directed to the SPIRAL identification station, 
where the experimental setup was mounted. The beam line to this detection station was regularly optimized by means of
a stable $^{40}$Ar$^{3+}$ beam produced with the ECR source. Secondary beam intensities of 1200 pps on average were obtained
for $^{33}$Ar$^{3+}$. The $^{33}$Ar beam was contaminated by small amounts of $^{35}$Ar possibly transmitted as a gas to 
the detection setup.

Beam profilers and Faraday cups close to the detection setup allowed to optimize the beam position in the experimental setup and
to monitor the beam intensity (mainly stable isotopes or molecules). The secondary beam was finally intercepted in the center
of the experimental setup by means of a 0.9~$\mu$m thick aluminized Mylar foil (1~cm $\times$ 2~cm) mounted on a thin metallic frame.

\begin{figure}
\begin{center}
\resizebox{.4\textwidth}{!}{\includegraphics{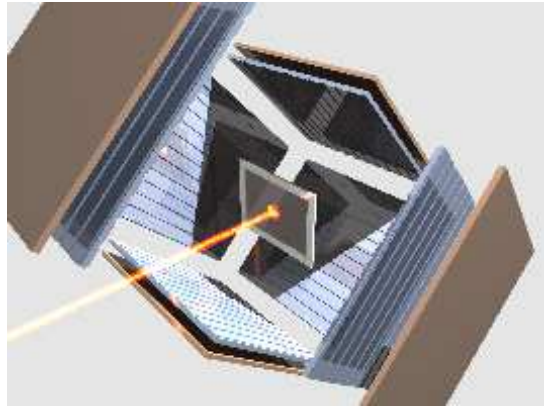}}
\caption[]{(Color online) Schematic view of the Silicon Cube consisting of 6 DSSSDs backed by 6 PIPS detectors. 
           The position of the mylar foil catcher and the entrance of the secondary beam are also shown.
	   }
\label{fig:setup}
\end{center}
\vspace*{-0.3cm}
\end{figure}

\begin{figure}
\begin{center}
\resizebox{.47\textwidth}{!}{\includegraphics{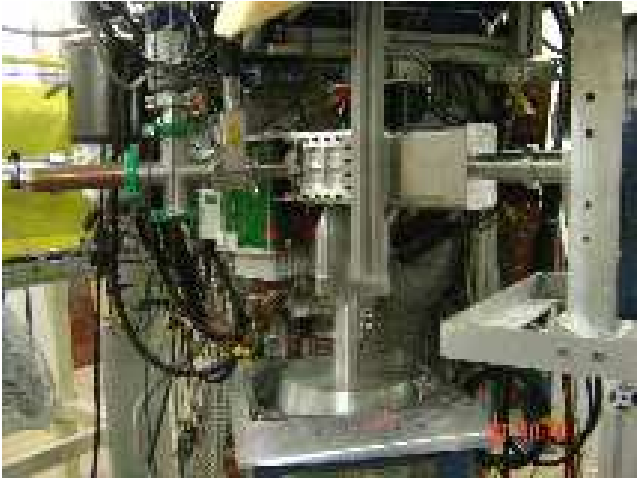}}
\caption[]{(Color online) Photo of the setup as installed at the low-energy identification station of SPIRAL at GANIL. The silicon cube 
           detector is seen in the center of the photo. On the left, the vacuum pipe to insert the catcher is visible. Two
           out of the three EXOGAM clover detectors are mounted and displayed on the bottom and right-hand side of the photo.
	   }
\label{fig:photo}
\end{center}
\vspace*{-0.3cm}
\end{figure}

The standard detection setup of the SPIRAL identification station was replaced by the Silicon Cube detector~\cite{matea09} 
and three high-efficiency germanium detectors from the EXOGAM array~\cite{exogam}. 
The Silicon Cube consists of six double-sided
silicon strip detectors (DSSSDs) with 16$\times$16 strips and a pitch of 3~mm forming a cube surrounding the mylar catcher foil. 
These DSSSDs were backed by six large-area (50x50~mm$^2$) Passivated Implanted Planar Silicon (PIPS) detectors used to detect
$\beta$ particles. The thicknesses of the different detectors are given in Table~\ref{tab:thickness}. The arrangement
of the different detectors can be seen on figure~\ref{fig:setup}. A photo of the setup installed at the low-energy
identification station of SPIRAL at GANIL is shown in figure~\ref{fig:photo}.

\begin{table}[hht]
\begin{center}
\begin{tabular}{lcccccc}
\hline
\hline
DSSSD - PIPS position &  1  &  2  &  3  &  4  &  5  &  6  \\
\hline
DSSSD thickness ($\mu$m) & 300 & 287 & 270 &  64	& 1000 & 288 \\
PIPS thickness ($\mu$m) & 300 & 300 & 500 & 1473& 150 & 1498 \\
\hline
\hline
\end{tabular}
\end{center}
\caption{Thicknesses of the different silicon detectors used in the present study. The arrangement of the detectors can be
         seen in figure~\ref{fig:setup}.}
\label{tab:thickness}
\end{table}

The 192 signal channels from the DSSSDs were readout via a printed circuit board incorporated in the detector housing.
The 16-channel pre-amplifier cards were mounted directly on the detector chamber enabling thus a rather compact configuration.
The six channels of the PIPS detectors were connected to their pre-amplifiers via standard LEMO connectors. This setup
reaches a geometrical efficiency for proton detection of 54\%~\cite{matea09}. 

The Silicon Cube was surrounded by three EXOGAM clover detectors. These detectors were calibrated in efficiency
and energy by means of standard calibration sources and yielded a total efficiency of about 3.2\% at 1332~keV.

Due to technical problems, DSSSD 2 did not function correctly and was excluded from the present analysis.
The thickness of detector 4 limited its use to low-energy protons. Therefore, it was also excluded from the analysis. 
Due to its large thickness, DSSSD 5 yields a quite high $\beta$ tail thus worsening its energy resolution. We used this 
detector only for the high-energy part of the spectrum (above 6~MeV) where the other detectors did no longer fully stop the protons. 
Therefore, we used detectors 1, 3, and 6 to determine proton energies and branching ratios for proton lines below 6~MeV and detector
5 above. The spectrum above 6~MeV was corrected for the geometrical efficiency of detector 5 with respect to detectors 1, 3, and 6.
The $\beta$ signal in the PIPS detectors was used in the analysis to reduce the $\beta$-p summing by requiring to have the $\beta$ signal 
in any of the other back detectors of the cube. This condition permitted to significantly reduce the $\beta$ tails in the proton spectra.

All silicon detectors were first calibrated by means of a triple $\alpha$ source ($^{239}$Pu, $^{241}$Am, $^{244}$Cm). The final
energy calibration of the DSSSDs was performed using the $\beta$-delayed proton peaks  to the ground state 
of $^{32}$S from the IAS recently remeasured~\cite{pyle02} to be at 5547.8(9) keV and the 3/2$^+$ state at 
3971.9(12) keV~\cite{endt90} excitation energy in $^{33}$Cl, respectively. An energy resolution of 50~keV
was obtained for the DSSSD sum spectrum from detectors 1, 3, and 6. Events in the DSSSDs were accepted only if the
signals in the front and back side were within $\pm$ 100~keV of each other.

Different runs optimized for $^{33}$Ar were performed yielding a total of 28~h for this isotope.
Several runs were also performed for $^{31}$Ar. These latter settings were strongly contaminated by $^{33}$Ar
and yielded rather low production rates for $^{31}$Ar. These data were thus only used to study the performance of the 
Silicon Cube~\cite{matea09}.

For the data taking with the different argon isotopes, only the DSSSDs were allowed to trigger the data acquisition.
After a trigger, all channels were readout by means of the GANIL data acquisition system via VXI and VME modules.

\section{Analysis procedure and experimental results}

\begin{figure}[hht]
\begin{center}
\resizebox{.32\textwidth}{!}{\includegraphics{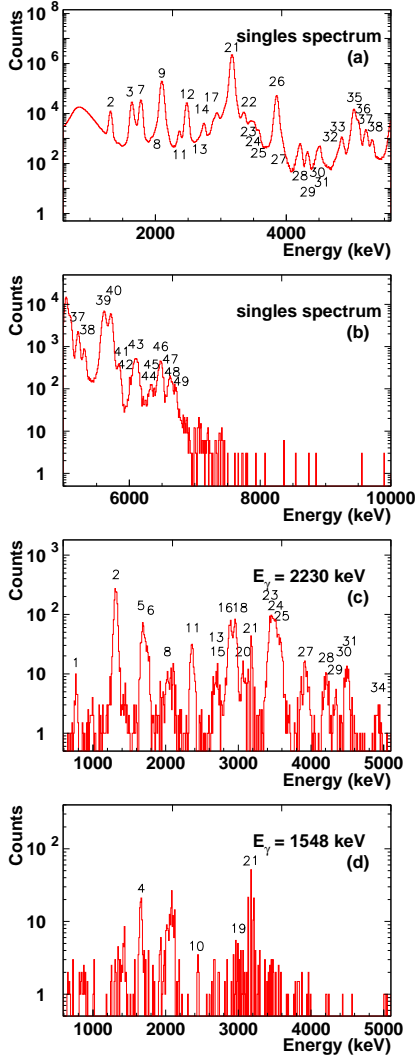}}
\caption[]{(Color online) (a,b) The figure shows the $\beta$-gated proton spectrum as obtained from the strips on the front side 
           of detectors 1, 3, 5, and 6. The peak numbers correspond to those of table~\ref{tab:protons}. Numbers below the curve are 
           for protons decaying to the first excited state, but visible in the unconditioned spectrum. 
           The decay from the IAS in $^{33}$Cl to the ground state of $^{32}$S is by far the most prominent peak.
           (c) The spectrum is generated from the same detectors, however, with
           a condition that the 2230 keV $\gamma$ ray is observed in the germanium detectors. (d) The spectrum is
           obtained in a similar manner with a condition of the 1548 keV $\gamma$ ray. No subtraction is operated in the sense that 
           spectrum (a,b) contains also proton groups decaying to the first excited state (e.g. the 1317~keV protons), whereas spectrum (c)
           contains also events in coincidence with the 1548~keV $\gamma$ line, as this 2230~keV $\gamma$ ray always follows the 1548~keV
           $\gamma$ ray. The area of the proton peak at 3173~keV observed in the coincidence spectra with the 2230 and 1548~keV $\gamma$
           lines gives the level of random coincidences of about 2$\times$10$^{-4}$. For spectra (c) and (d) we indicate the 
           reminder of this 3173~keV peak number 21.
	   }
\label{fig:protons}
\end{center}
\vspace*{-0.3cm}
\end{figure}

The experimental data were analyzed as singles and coincidence data. The different peaks were fitted and integrated by means
of the PAW package from the CERN library. Most of these peaks are due to $\beta$-delayed proton emission from states
in the $\beta$-decay daughter nucleus $^{33}$Cl to the ground states of $^{32}$S. Decays to excited
states were identified by means of proton-$\gamma$ coincidences with the 2230~keV and the 1548~keV $\gamma$ lines 
from the decay of excited states in $^{32}$S.
Figure~\ref{fig:protons} shows the $^{33}$Ar $\beta$-gated proton spectrum and the proton spectra coincident with the $\gamma$ lines at
2230~keV and the 1548~keV, respectively. 
 
\begin{figure}
\begin{center}
\resizebox{.38\textwidth}{!}{\includegraphics{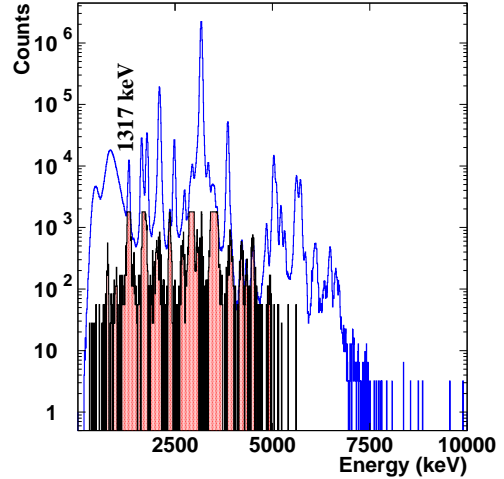}}
\caption[]{(Color online) The figure shows the $\beta$-gated proton spectrum as obtained from the strips on the front side 
           of detectors 1, 3, 5, and 6 overlaid with the spectrum obtained from the 2230~keV $\gamma$-ray coincidence spectrum.
           The normalization of the second spectrum is such that the intensities of the 1317~keV peak, being to 100\% due
           to the 2230~keV $\gamma$ coincidence, match. To generate the $\gamma$-coincident spectrum, a spectrum was first generated
           with a 20~keV wide condition on the 2230~keV $\gamma$ line. From this spectrum, we subtract then a similar spectrum
           conditioned by a gate left and right of the $\gamma$ line.
	   }
\label{fig:prot_gam}
\end{center}
\vspace*{-0.3cm}
\end{figure}

At several instances, peaks due to proton emission to the ground state and to the excited states overlay. To ensure correct
identification due to the different decay branches, the $\gamma$-gated proton spectrum for the 2230~keV $\gamma$ line was normalized 
to the $\beta$-gated spectrum by means of the proton peak at 1317~keV which is assumed to be only due to emission to the first
excited state~\cite{schardt93}. As this line is seen in the $\beta$-gated spectrum and, of course, in the spectrum gated with
the 2230~keV $\gamma$ line, both spectra can be normalized with respect to each other and a common relative normalization
can be established. A fit of the 1317~keV peak in both spectra, $\gamma$ gated and singles, yielded a factor of 56 with which the
$\gamma$-gated spectrum has to be multiplied to match both spectra. As the error of this number does not influence the final result 
in any significant way, it was not included. Figure~\ref{fig:protons}c and the shaded part of 
figure~\ref{fig:prot_gam} show this spectrum dominated by this proton peak.

For the proton spectrum gated by the 1548~keV $\gamma$ line, we normalized the proton spectrum with the same factor as
the spectrum in coincidence with the 2230~keV line and the ratio of
the $\gamma$ detection efficiencies of the 1548~keV line and the 2230~keV line. This exploits the idea that in such an analysis the
probability to observe protons in coincidence with the 1548~keV line is only modulated with respect to the spectrum gated by
the 2230~keV line by the $\gamma$-detection efficiency. The ratio of the two $\gamma$-ray efficiencies was 
$\epsilon_{1548}$/$\epsilon_{2230}$~= 2.44\%/1.89\%. Thus, the spectrum generated with a $\gamma$ coincidence of the 1548~keV line 
was multiplied by this factor and the factor determined between the $\beta$-gated spectrum and the spectrum produced with 
the 2230~keV coincidence in order to match the relative proton intensities for the ground state 
transitions, the transitions to the 2230~keV state and to the 3778~keV state in $^{32}$S. 

\clearpage

To determine proton and $\gamma$-ray intensities, we used Gaussians for the peaks and straight-line fits for the background. All 
intensities were first normalized with respect to the most intense peak (the 3173~keV proton line and the 810~keV $\gamma$ line, 
respectively) to obtain relative branching ratios.

The relative intensities of the proton groups can easily be converted into absolute intensities by assuming that the absolute intensity
of the feeding of the IAS at 5549~keV can be reliably determined. For this purpose, one usually assumes that the Fermi strength can be
calculated more or less model independently by neglecting Coulomb and radiative corrections (see e.g.~\cite{hardy09}). In the
case of $^{33}$Ar as for other odd-mass nuclei, the additional problem of a GT contribution to the decay to the IAS arises.
This contribution can either be determined experimentally (see e.g. \cite{schardt93,garcia00}), be neglected assuming its 
contribution to be small, or taken from theoretical predictions. We will calculate the Fermi part neglecting the corrections 
mentioned above and take the GT contribution from Ref.~\cite{garcia00}.

The Fermi strength can be calculated from the relation between the $ft$ value and the Fermi matrix element $M_F$: 
$ ft = K / (g$'$_v^2 * M_F^2)$, where K is a constant and $g$'$_v$ is the effective vector coupling constant of the weak interaction.
For $K / g$'$_v^2$, we use a value of 6144.2(13)~s deduced from the super-allowed 0$^+ \rightarrow$ 0$^+$ decays~\cite{hardy09}. 
The matrix element for nuclei with an isospin quantum number $T = 3/2$ is $\sqrt{3}$.
This allows the Fermi $ft$ value to be calculated to be 2048~s.

\renewcommand{\baselinestretch}{0.}
\begin{table*}[ht]
\vspace*{-0.5cm}
\begin{center}
\mbox{\footnotesize
\begin{tabular}{cccccc}
\hline
\hline
Peak   &  Energy     &  BR (\%)  &   BR (\%)                       &        BR (\%)              & BR (\%) \\
number &  (keV)      & this work & Schardt et al.~\cite{schardt93} & Borge et al.~\cite{borge87} & average \\
\hline
   \multicolumn{6}{l}{protons to the ground state of $^{32}$S}  \\
3 &  1645(2)        &  0.411(20)    &  0.391(6)                &   0.343(10)   &  0.380(16)     \\
7 &  1781(2)        &  0.471(22)    &  0.459(6)                &   0.434(10)   &  0.453(8)      \\
9 &  2100(3)        &  2.73(12)     &2.37(2)+0.35(1)$\;^{(*)}$ &   2.37(2)     &  2.54(12)      \\
12&  2481(2)        &  0.362(17)    &  0.353(6)                &   0.333(10)   &  0.349(7)      \\
14&  2744(3)        &  0.0483(44)   &  0.0403(31)              &   0.0454(50)  &  0.0435(25)    \\
17&  2941(4)        &  0.0748(55)   &  0.0713(31)              &   0.1222(81)  &  0.0772(107)   \\
21&  3173(3)        & 31.0(14)      & 31.0(14)                 &  31.0(14)     & 31.0(14)       \\
22&  3350(4)        &  0.0918(48)   &  0.0310(62)              &   0.7573(404) &  0.0753(496)   \\
26&  3857(3)        &  0.735(34)    &  0.716(6)                &   0.808(20)   &  0.724(18)     \\
32&  4719(5)        &  0.00079(10)  & -                        &          -    &  0.00079(10)   \\
33&  4860(4)        &  0.0097(8)    &  0.0152(31)              &   0.0232(30)  &  0.0108(24)    \\
35&  5039(4)        &  0.224(12)    &  0.217(6)                &   0.333(10)   &  0.245(34)     \\
36&  5101(4)        &  0.0470(50)   &  0.0589(31)              &   0.1111(101) &  0.0592(103)   \\
37&  5225(4)        &  0.0234(24)   &  0.0288(16)              &   0.0545(40)  &  0.0298(59)    \\
38&  5317(3)        &  0.0083(12)   &  0.0133(12)              &   0.0182(30)  &  0.0113(22)    \\
39&  5623(3)        &  0.124(7)     &  0.118(25)               &   0.222(10)   &  0.155(32)     \\
40&  5723(3)        &  0.092(5)     &  0.062(15)               &   0.172(10)   &  0.103(22)     \\
41&  5855(9)        &  0.00284(40)  &     -                    &   0.01212(202)&  0.0032(18)    \\
42&  6011(10)       &  0.00100(15)  &     -                    &         -     &  0.00100(15)   \\
43&  6100(10)       &  0.0138(15)   &     -                    &   0.0212(40)  &  0.0147(24)    \\
44&  6344(8)        &  0.00053(9)   &     -                    &   0.00505(202)&  0.00054(21)   \\
45&  6389(10)       &  0.00027(8)   &     -                    &         -     &  0.00027(8)    \\
46&  6480(10)       &  0.0103(11)   &     -                    &   0.0162(20)  &  0.0116(24)    \\
47&  6628(10)       &  0.00170(23)  &     -                    &   0.00404(202)&  0.00173(26)   \\
48&  6657(9)        &  0.00049(10)  &     -                    &         -     &  0.00049(10)   \\
49&  6715(9)        &  0.00100(12)  &     -                    &   0.00364(162)&  0.00102(20)   \\
  &  6800 - 6900    &  0.00023(9)   &     -                    &         -     &  0.00023(9)    \\
  &  6900 - 7000    &  0.00007(3)   &     -                    &         -     &  0.00007(3)    \\
  &  7000 - 7100    &  0.00032(4)   &     -                    &         -     &  0.00032(4)    \\
  &  7100 - 7200    &  0.00000(0)   &     -                    &         -     &  0.00000(0)    \\
  &  7200 - 7300    &  0.00012(3)   &     -                    &         -     &  0.00012(3)    \\
  &  7300 - 7400    &  0.00010(3)   &     -                    &         -     &  0.00010(3)    \\
  &  7400 - 7500    &  0.00008(3)   &     -                    &         -     &  0.00008(3)    \\
  &  7500 - 8000    &  0.00006(3)   &     -                    &         -     &  0.00006(3)    \\
  &  8000 - 9000    &  0.00004(3)   &     -                    &         -     &  0.00004(3)    \\
\hline
  \multicolumn{6}{l}{protons to the first excited state of $^{32}$S}  \\
1 &   762(10)       &  0.0202(17)   &      -               &         -       &  0.0202(17)     \\
2 &  1317(8)        &  0.168(9)     &   0.180(3)           &    0.191(8)     &  0.180(4)       \\
3 &  1691(6)        &  0.0332(32)   &   0.0319(16)         &    0.0464(61)   &  0.0329(22)     \\
6 &  1764(5)        &  0.0081(13)   &   0.0220(31)         &       -         &  0.0102(49)     \\
8 &  2024(5)        &  0.0043(7)    &      -               &       -         &  0.00428(71)    \\
11&  2370(5)        &  0.0153(12)   &   0.0158(31)         &    0.0192(30)   &  0.0159(9)      \\
13&  2710(7)        &  0.0069(12)   &      -               &       -         &  0.0069(12)     \\
15&  2810(10)       &  0.00141(14)  &      -               &       -         &  0.00141(14)    \\
16&  2886(7)        &  0.0376(35)   &   0.0341(31)         &    0.0646(61)   &  0.0394(70)     \\
18&  2957(7)        &  0.0359(32)   &   0.0434(31)         &       -         &  0.0397(38)     \\
20&  3066(6)        &  0.00440(61)  &       -              &    0.7068(2020) &  0.0045(20)     \\
23&  3469(6)        &  0.0531(40)   &       -              &    0.4140(404)  &  0.057(36)      \\
24&  3515(6)        &  0.0150(25)   &  0.0065(16)          &       -         &  0.0089(38)     \\
25&  3576(5)        &  0.0085(16)   &       -              &    0.1010(101)  &  0.011(14)      \\
27&  3926(5)        &  0.0082(13)   &       -              &    0.0192(40)   &  0.0091(31)     \\
28&  4209(5)        &  0.00645(83)  &       -              &    0.0172(30)   &  0.0072(27)     \\
29&  4330(8)        &  0.00142(40)  &       -              &    0.0081(20)   &  0.0017(13)     \\
30&  4474(5)        &  0.00367(55)  &       -              &         -       &  0.00367(55)    \\
31&  4505(6)        &  0.00467(62)  &       -              &    0.01820(51)  &  0.0049(16)     \\
34&  4923(6)        &  0.00066(16)  &       -              &         -       &  0.00066(16)    \\
\hline
 \multicolumn{6}{l}{protons to the second excited state of $^{32}$S}  \\
4 &  1665(6)        &  0.0060(11)   &  0.0099(16) &                    -      &  0.0074(19)    \\
10&  2368(6)        &  0.0012(3)    &      -      &                    -      &  0.0012(3)     \\
19&  3016(10)       &  0.0007(2)    &      -      &                    -      &  0.0007(2)     \\
\hline
\hline
\end{tabular}
}
\end{center}
\vspace*{-0.5cm}
\renewcommand{\baselinestretch}{1}
\caption{Summary of proton energies in the laboratory frame (from this work) and absolute branching ratios from the present work, from 
         Schardt et al.~\cite{schardt93} and from Borge et al.~\cite{borge87}. Eight proton lines attributed by Borge et al. to a 
         transition to the ground state were newly attributed to proton emission to the first excited state. One proton group of 
         Schardt et al. is attributed to the decay to the second excited state. $^{(*)}$ Two proton lines observed by Schardt et 
         al.~\cite{schardt93} are neither resolved in the present work nor in the work of Borge et al.~\cite{borge87}. The absolute 
         intensities have been determined by normalizing all proton group with respect to the 3173~keV line for which the absolute 
         branching ratio was established as described in the text. The last column gives the error-weighted averages of the previous 
         three columns. }
\label{tab:protons}
\end{table*}

\clearpage

We take the GT to Fermi ratio from Ref.~\cite{garcia00}, where
it was determined to be 0.044$\pm$0.002. Adding this value to the Fermi matrix element and taking into account the ratio between the
squares of the vector and the axial-vector coupling constants of 1.2695(29)~\cite{pdg08}, we derive a total $ft$ value for the decay 
to the IAS of 1962(5)~s. As we have neglected the corrections mentioned above, we add an additional uncertainty of 1\% for this number
which yields the final $ft$ value for the IAS of 1962(20)~s.

\begin{table}[tth]
\begin{center}
\begin{tabular}{crcrcc}
\hline
\hline
          & \multicolumn{2}{c}{this work} & \multicolumn{2}{c}{Borge et al.~\cite{borge87}} & average \\
          & E$_\gamma$  (keV) & BR (\%) &  E$_\gamma$  (keV) & BR (\%) & BR (\%) \\
\hline
$^{33}$Cl & 810.6(2)          & 100(1)  &  810.3(5)          & 100(3)  &  100     \\
$^{33}$Cl & 1541.4(6)         & 3.6(2)  & 1541.5(5)          & 2.4(5)  & 3.4(4)  \\
$^{32}$S  & 2230.3(5)         & 6.1(2)$^*$  & 2230.6(9)          & 1.7(5)  & -       \\
$^{33}$Cl & 2352.5(6)         & 1.3(2)  & 2352.2(9)          & 1.7(5)  & 1.4(2)  \\
$^{33}$S  & 2866.2(9)         & 0.6(1)  & 2867.7(9)          & 1.3(1)  & -  \\
\hline
\hline
\end{tabular}
\end{center}
\caption{Gamma-ray energies and their relative branching ratios from the present work and the work of Borge et al.~\cite{borge87}.
          Averages are given only for $\gamma$ rays following the decay of $^{33}$Ar. For the $\gamma$ ray at 2230~keV, the relative
          branching ratio in our experiment indicated by the star is too high, when it is calculated with respect to the 810~keV line. 
          The former $\gamma$ ray is always preceded by a proton which triggers the DSSSDs, while for the latter the $\beta$ 
          particles have to trigger. Evidently, the $\beta$ trigger probability is much smaller than the proton trigger efficiency 
          which then yields higher relative branching ratios for $\beta$p$\gamma$ events. The literature value for the last 
          $\gamma$-ray energy is 2867.72(2)~keV.}
\label{tab:gammas}
\end{table}

To determine the absolute branching ratio to the IAS, we now need the statistical rate function or Fermi function calculated from the
$\beta$-decay Q$_{EC}$ value (11.6193(6)~MeV~\cite{audi03}) from which we subtract the excitation energy of the IAS (5.5478(9)~MeV)~\cite{pyle02}
and the half-life of $^{33}$Ar. For the half-life, we use the average of 
different literature values~\cite{reeder64,hardy64,poskanzer66,hardy71,borge87} for which we find 173.9(9)~ms. The statistical
rate function is calculated to be 3501(4). With these numbers, we determine a branching ratio to the IAS of 31.0(14)\%.
 
The branching ratio to the ground state of $^{33}$Cl could not be determined in our experiment. We therefore assume that mirror
symmetry is a good approximation and calculate its feeding from the $ft$ value of the mirror decay. $^{33}$P decays with a 100\%
branch to the ground state of $^{33}$S. The half-life (25.34(12)~d) and the Q$_EC$ value (248.5(11)~keV)~\cite{audi03} enables us 
to determine the $ft$ value to be 107280(2250)~s ($logft$ = 5.03(1)). Using this $ft$ value for the decay to the ground state 
of $^{33}$Cl, we determine a branching ratio of 18.7(4)\%.

\begin{figure}
\begin{center}
\resizebox{.49\textwidth}{!}{\includegraphics{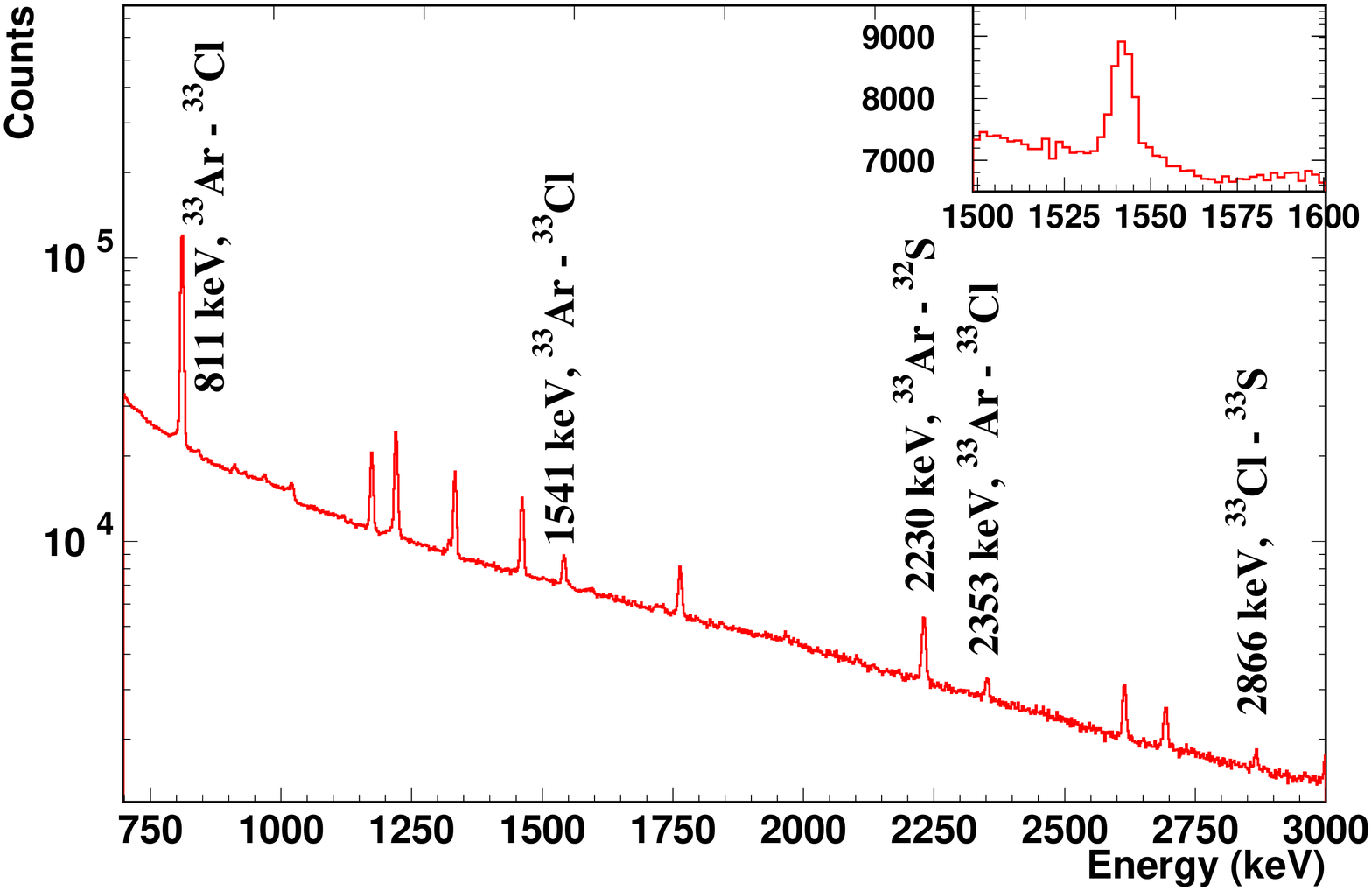}}
\caption[]{(Color online) Gamma-ray spectrum registered with the three germanium clover detectors. The peaks labeled belong directly or 
           indirectly to the decay of $^{33}$Ar. The inset shows the region around the 1541~keV and 1548~keV $\gamma$ rays.
           The 1548~keV peak from the decay of the second excited state in $^{32}$S is very weak and only visible as a shoulder 
           of the stronger 1541~keV peak from the decay of the second excited state of $^{33}$Cl. The relative intensities 
           of all relevant peaks can be found in table~\ref{tab:gammas}. Background lines are from $^{60}$Co 
           (1173~keV, 1332~keV), from a contamination of the beam with $^{35}$Ar (1220~keV, 1764~keV, 2694~keV), from $^{40}$K (1461~keV), 
           and from the decay of the first excited state of $^{208}$Pb (2615~keV).
	   }
\label{fig:gammas}
\end{center}
\end{figure}

The difficulty is now to convert the branching ratios for the feeding of bound levels in $^{33}$Cl to absolute branching ratios.
We will test two ways of determining these branching ratios. In the first procedure, we use the total proton branching ratio
and the branching ratio to the ground state to deduce the feeding of bound excited levels as the three contributions should 
add up to 100\%, as long as $\gamma$ decay of proton-unbound levels is negligible. In the second, we use the decay of $^{33}$Cl 
and the relative intensities of the $\gamma$ transition in $^{33}$Cl to determine the feeding of bound levels.

For the first procedure, we can obtain the total proton branching ratio in two ways: i) by summing all proton branching ratios in 
table~\ref{tab:protons}, ii) by integrating the $\beta$-gated proton spectrum assuming that it contains all proton events.
However, in second this case, one has to cut the $\beta$ particles by means of a threshold which we chose, somewhat arbitrarily, to be 1000~keV. 
We note that the final result changes by 0.8\%, if we integrate the proton spectra starting at 900 keV. We include this in the final error. 
The total proton rate can then be normalized by means of the known absolute branching ratio of the IAS of 31.0(14)\%. We 
determine a ratio between the number of counts for the IAS proton line and the total proton intensity of 78(2)\%. In the former
procedure, where we sum the branching ratios from table~\ref{tab:protons}, we obtain a total proton branching ratio of 36.9(17)\%.
The latter procedure gives a value of 39.7(12)\%. We adopt a final value of 38.8(13)\% for this first procedure.

With the branching ratio to the ground state of $^{33}$Cl of 18.7(4)\% and the total proton branching ratio of 38.8(13)\%, 
we derive a branching ratio to bound excited states in $^{33}$Cl of 42.5(17)\%. From this number and the relative $\gamma$ 
branching ratios (see figure~\ref{fig:gammas} and table~\ref{tab:gammas}), one can determine the absolute branching ratios 
for the feeding of the 810~keV level of 40.5(16) \% and for the 2352~keV level of 2.0(3)\%.

\renewcommand{\baselinestretch}{0.}
\begin{table*}[hht]
\vspace*{-1.3cm}
\begin{center}
\mbox{\footnotesize
\begin{tabular}{ccccccccccc}
\hline
\hline
\multicolumn{1}{c}{E$_x$ (keV) } & spin/parity  &\multicolumn{1}{c}{E$_x$ (keV)} & 
\multicolumn{1}{c}{E$_\gamma$ (keV)} & BR (\%) & B(GT) & $\Sigma$B(GT) & & \multicolumn{2}{c}{USD shell model~\cite{brown09}}\\
\cite{endt98}&\cite{endt98}&(this work)&&&&&& E$_x$ (keV) & BR (\%) \\
\hline
   0      &  3/2+        & 0        &     0       & 18.7  &  0.0363 &  0.0363  & & 0     & 13.7404  \\
 810.5(2) &  1/2+        & 810.6(2) &   810.6     & 40.5  &  0.1151 &  0.1514  & & 779   & 49.0350  \\
2352.5(4) &  3/2+        &2352.5(6) &  2352.5     &  1.40 &  0.0128 &  0.1642  & & 2174  &  5.5445  \\
          &              &          &  1541.4     &  0.60 &         &          & &       &          \\
[0.5em]
\hline
\hline
\multicolumn{1}{c}{E$_x$ (keV) } & spin/parity &\multicolumn{1}{c}{E$_x$ (keV)} & 
\multicolumn{1}{c}{E$_p$ (keV)} & BR (\%) & B(GT) & $\Sigma$B(GT) & spin/parity &\multicolumn{2}{c}{USD shell model~\cite{brown09}}\\
\cite{endt98}&\cite{endt98}&(this work)&&&&& (this work) & E$_x$ (keV) & BR (\%) \\
\hline
3971.9(12) &  3/2+       & 3973(2)  &  1645       &   0.380   &  0.0069  &  0.1711  & & 3667  &  0.3296  \\
4112.8(8)  &  (1/2,3/2)+ & 4113(2)  &  1781       &   0.453   &  0.0090  &  0.1801  & & 4185  &  0.3313  \\
4438.2(15) &  1/2+       & 4442(3)  &  2100       &   2.54    &  0.0646  &  0.2447  & & 4286  &  2.9158  \\
4832(2)    &  3/2+       & 4835(2)  &  2481       &   0.349   &  0.0121  &  0.2568  & & 4808  &  0.2482  \\
5104(2)    &  3/2+       & 5106(3)  &  2744       &   0.0435  &  0.0019  &  0.2587  & & 5076  &  0.2798  \\
           &             & 5310(4)  &  2941       &   0.0772  &  0.0040  &  0.2627  & (3/2)+ &        &          \\
           &             &          &   762$\#$   &   0.0202  &  0.0010  &  0.2637  & &        &          \\
5544(1)    &  1/2+       & 5549(3)  &  3173       &  31.0     &  0.1320  &  0.3957  & & 5191  & 25.7003  \\
5731(3)    &  1/2+       & 5731(4)  &  3350       &   0.0753  &  0.0057  &  0.4014  & &        &          \\
           &             & 5865(8)  &  1317$\#$   &   0.180   &  0.0155  &  0.4169  & (3/2)+ &        &          \\
6248(3)    &  (1/2,3/2)+ & 6253(3)  &  3857       &   0.724   &  0.0927  &  0.5097  & &        &          \\
or 6257(5) &             &          &  1691$\#$   &   0.0329  &  0.0042  &  0.5139  & &        &          \\
           &             & 6326(5)  &  1764$\#$   &   0.0102  &  0.0014  &  0.5153  & (3/2)+ &        &          \\
           &             & 6594(5)  &  2024$\#$   &   0.00428 &  0.0008  &  0.5161  & (3/2)+ &        &          \\
           &             & 6951(5)  &  2370$\#$   &   0.0159  &  0.0045  &  0.5206  & (3/2)+ &        &          \\
           &             & 7143(5)  &  4719       &   0.00079 &  0.0003  &  0.5209  & &        &          \\
7272(4)    &  3/2+       & 7292(5)  &  4860       &   0.0108  &  0.0048  &  0.5257  & (3/2)+ &        &          \\
           &             &          &  2710$\#$   &   0.00694 &  0.0031  &  0.5288  & &        &          \\
           &             & 7405(10) &  2810$\#$   &   0.00141 &  0.0007  &  0.5295  & (3/2)+ &        &          \\
           &             & 7475(4)  &  5039       &   0.245   &  0.1398  &  0.6693  & &        &          \\
           &             &          &  2886$\#$   &   0.0394  &  0.0228  &  0.6921  & &        &          \\
           &             & 7537(4)  &  5101       &   0.0592  &  0.0371  &  0.7292  & &        &          \\
           &             & 7556(7)  &  2957$\#$   &   0.0397  &  0.0256  &  0.7548  & (3/2)+ &        &          \\
           &             & 7666(4)  &  5225       &   0.0298  &  0.0226  &  0.7773  & &        &          \\
           &             &          &  3066$\#$   &   0.00446 &  0.0034  &  0.7807  & &        &          \\
           &             & 7762(5)  &  5317       &   0.0113  &  0.0099  &  0.7907  & (1/2)+ &        &          \\
           &             &          &  1665$\#\#$ &   0.00735 &  0.0066  &  0.7972  & &        &          \\
           &             & 8077(4)  &  5623       &   0.155   &  0.2295  &  1.0268  & &        &          \\
           &             &          &  3469$\#$   &   0.0567  &  0.0853  &  1.1121  & &        &          \\
           &             & 8132(6)  &  3515$\#$   &   0.00886 &  0.0145  &  1.1266  & (3/2)+ &        &          \\
           &             & 8182(7)  &  5723       &   0.103   &  0.1833  &  1.3099  & &        &          \\
           &             &          &  3576$\#$   &   0.0108  &  0.0198  &  1.3297  & &        &          \\
           &             & 8315(9)  &  5855       &   0.00318 &  0.0073  &  1.3370  & &        &          \\
           &             & 8491(9)  &  6011       &   0.00100 &  0.0032  &  1.3401  & (1/2)+ &        &          \\
           &             &          &  2368$\#\#$ &   0.00117 &  0.0039  &  1.3440  & &        &          \\
           &             & 8557(5)  &  6100       &   0.0147  &  0.0562  &  1.4003  & &        &          \\
           &             &          &  3926$\#$   &   0.00911 &  0.0349  &  1.4352  & &        &          \\
           &             & 8819(8)  &  6344       &   0.00054 &  0.0037  &  1.4389  & &        &          \\
           &             & 8847(5)  &  4209$\#$   &   0.00719 &  0.0523  &  1.4911  & (3/2)+ &        &          \\
           &             & 8865(10) &  6389       &   0.00027 &  0.0021  &  1.4932  & &        &          \\
           &             & 8967(6)  &  6480       &   0.0116  &  0.1122  &  1.6054  & (1/2)+ &        &          \\
           &             &          &  4330$\#$   &   0.00167 &  0.0166  &  1.6221  & &        &          \\
           &             & 9119(5)  &  6628       &   0.00173 &  0.0252  &  1.6473  & &        &          \\
           &             &          &  4474$\#$   &   0.00367 &  0.0549  &  1.7022  & &        &          \\
           &             & 9152(6)  &  6657       &   0.00049 &  0.0077  &  1.7099  & (3/2)+ &        &          \\
           &             &          &  4505$\#$   &   0.00487 &  0.0798  &  1.7897  & &        &          \\
           &             &          &  3016$\#\#$ &   0.00066 &  0.0113  &  1.8010  & &        &          \\
           &             & 9202(9)  &  6715       &   0.00102 &  0.0193  &  1.8202  & &        &          \\
           &             &9300-9400 &  6800-6900  &   0.00023 &  0.0068  &  1.8270  & &        &          \\
           &             &9400-9500 &  6900-7000  &   0.00007 &  0.0030  &  1.8300  & &        &          \\
           &             &9500-9600 &  7000-7100  &   0.00032 &  0.0194  &  1.8494  & &        &          \\
           &             &9584(6)   &  4923$\#$   &   0.00066 &  0.0459  &  1.8954  & (3/2)+ &        &          \\
           &             &9600-9700 &  7100-7200  &   0.00000 &  0.0000  &  1.8954  & &        &          \\
           &             &9700-9800 &  7200-7300  &   0.00012 &  0.0175  &  1.9129  & &        &          \\
           &             &9800-9900 &  7300-7400  &   0.00010 &  0.0221  &  1.9350  & &        &          \\
           &             &9900-10000&  7400-7500  &   0.00008 &  0.0315  &  1.9666  & &        &          \\
           &            &10000-10500&  7500-8000  &   0.00006 &  0.1042  &  2.0708  & &        &          \\
           &            &10500-11500&  8000-9000  &   0.00004 &  0.4966  &  2.5674  & &        &          \\
[0.5em]                                                             
\hline
\hline
\end{tabular}
}
\end{center}
\renewcommand{\baselinestretch}{1}
\vspace*{-0.6cm}
\caption{Excitation energies, $\gamma$ and laboratory proton energies, branching ratios (averages from the present work, Borge et 
         al.~\cite{borge87}, and Schardt et al.~\cite{schardt93}), individual B(GT) values and the summed B(GT). In cases, where 
         the excitation energy is missing for the present work, the proton is emitted from the same level as the proton in the 
         previous line. Proton energies labeled with $\#$ are due to protons populating the first excited state, whereas proton 
         energies labeled with $\#\#$ are due to protons populating the second excited state. Shell-model values 
         for the excitation energies and the branching ratios are given for levels below the IAS. They were obtained using the USD
         effective interaction~\cite{brown88}. The excitation energies determined 
         in the present work are also compared to values from Endt~\cite{endt98}, if the level was not only determined from 
         $\beta$-decay studies. We used the same criteria to present the spin/parity values of column 2 from Ref.~\cite{endt98}. 
         Column 8 gives the spin values proposed from the present work (see text for discussion).}
\label{tab:bgt}
\end{table*}

\clearpage

In the second procedure to determine the total branching ratio for decays to bound states in $^{33}$Cl, we calculate these branching
ratios from the observation of the decay of $^{33}$Cl. The $\gamma$ ray at 2868~keV from this decay has a branching ratio 
of 0.44(6)\%~\cite{wilson80}. If we use the observed relative branching ratio for this $\gamma$ ray from table~\ref{tab:gammas}
and the possible branches populating the ground state of $^{33}$Cl, we determine a branching ratio of the 810~keV $\gamma$ ray 
of $^{33}$Cl of 51.4(120)\% and of 2.6(6)\% for the branching ratio to the 2352~keV level. These values agree with those
determined with the first procedure, but they are significantly less precise. We therefore adopt the values obtained with 
the first procedure.

The procedures just laid out assume that there is no $\gamma$ decay of proton decaying levels. In addition, forbidden transitions
could falsify the GT strength distribution calculated in the following. The possible presence of forbidden transitions is 
shortly discussed below.

With these absolute branching ratios for the feeding of the ground state of $^{33}$Cl as well as of the bound and unbound excited states 
of $^{33}$Cl, one can determine the feeding of all states in  the decay of $^{33}$Ar, the Gamow-Teller strength B(GT) and its sum.
These data are given in Table~\ref{tab:bgt}. For the 762~keV and 2941~keV proton lines, we adopted the level energy calculated 
only from the 2941~keV protons, as the low-energy protons are much stronger affected by the energy loss in the silicon dead layer 
and in the mylar foil of the catcher than the high-energy protons. For all other decays, we use average values from decays to the 
ground and first (second) excited states. In addition, we determine the log(ft) values for all observed transitions which are 
given in figure~\ref{fig:scheme}.

\section{Discussion of the results}

\subsection{Comparison with previous experimental work}

In the work of Borge et al.~\cite{borge87}, the energy calibration of the silicon detector at higher energies suffers from the 
fact that the calibration  was performed with $\alpha$ particles and that the pulse height defect~\cite{lennard86} was not 
corrected for. Therefore, with increasing energy above the IAS, the proton energies determined become smaller and smaller. 
In order to find the proton lines in Borge et al. corresponding to proton lines of the present work, we corrected for this 
effect of the Borge et al. data.
         
Compared to previous work~\cite{borge87,schardt93,honkanen96}, we observe all the proton lines identified in the work of Borge et al.~\cite{borge87}. 
However, eight of them are now attributed to proton emission to the first excited state rather than to the ground state (see 
table~\ref{tab:protons}). Generally, the branching ratios of Borge et al. are higher than our values.
As for Schardt et al.~\cite{schardt93}, the proton group at 1750~keV corresponds most likely to 
our proton energy of 1764(5)~keV observed in coincidence with the 2230~keV $\gamma$ ray.
The lines at 2121~keV and 2096~keV observed by Schardt et al. are not resolved in the present work. They correspond to
our line at 2100(3)~keV. The 1665~keV line is assigned to the decay to the second excited state, as observed in the present work. 
Honkanen et al.~\cite{honkanen96} identified two weak lines at 3485~keV and 5658~keV. The first one corresponds
probably to our line at 3469(6)~keV which we observe in coincidence with the $\gamma$ ray at 2230~keV (see below). For the higher-energy
line, we can not exclude that it corresponds to our proton line at 5623(3)~keV.

Table~\ref{tab:protons} gives the proton group energies, their absolute intensities from the different experiments and the average 
intensities. The branching ratios of the previous work were updated with respect to the branching ratio to the IAS as determined
above. For the experimental proton energies, we give only our energies. 

In table~\ref{tab:bgt}, we compare the levels observed in the present work with those found in the latest review of properties
of the mass A=33 chain by Endt~\cite{endt98}. The levels given in this compilation come both from $\beta$-decay studies~\cite{borge87} 
and from reaction work~\cite{bini72,aleonard76,wampfler80,wilkerson92}. For levels determined by different experimental approaches, 
we find reasonable agreement. For higher-energy levels, we just compare with those deduced from reaction work as the comparison 
with $\beta$-decay work has already been done in table~\ref{tab:protons}. Therefore, in table~\ref{tab:bgt}, we only compare with
previously established states deduced from work other than $\beta$ decay. Similarly, the spin values shown come only from reaction 
studies.

\subsection{Comparison with shell-model calculations and the GT strength distribution}

If we compare our results to shell-model calculations~\cite{brown09} (see table~\ref{tab:bgt}), we find that below the IAS 
only one predicted state at 3.863~MeV with a branching ratio of 0.0747\% is not observed experimentally. All other predicted 
states are observed experimentally with branching ratios in most cases close to the predicted ones.

\begin{figure}
\begin{center}
\resizebox{.42\textwidth}{!}{\includegraphics[angle=-90]{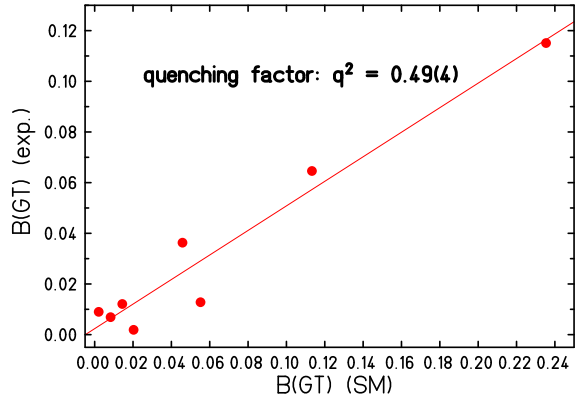}}
\caption[]{(Color online) Comparison of the GT strength between experiment and shell-model calculations for eight individual transitions of $^{33}$Ar 
           for which a correspondence between the shell-model calculations with the USD interaction and experiment was established in 
           table~\ref{tab:bgt}. The fit yields a quenching factor of 0.49(4). 
	   }
\label{fig:quenching}
\end{center}
\vspace*{-0.8cm}
\end{figure}

For a comparison of the experimental GT strength distribution with theoretical predictions, the theoretical strength has to be 
quenched. The origin of this quenching is not completely clear, but excitations of sub-nucleonic degrees of freedom, notably 
of the $\Delta$ resonance, and, probably to a much larger degree, contributions from outside the considered model space are 
most often suspected (see e.g.~\cite{arima99}). A typical quenching factor for the $sd$ shell is q$^2$ = 0.5. With our data, 
we can check this quenching factor. We will use experimental B(GT) values deduced for individual transitions below the IAS 
and compare them to the B(GT) strength for the equivalent transitions (see table~\ref{tab:bgt}) as predicted by theory. 
This comparison is shown in figure~\ref{fig:quenching}. A fit of these theoretical data as a function of the experimental 
data yields a quenching factor of 0.49(4). This is in perfect agreement with the generally accepted quenching in the $sd$ shell.

\begin{figure}
\begin{center}
\resizebox{.44\textwidth}{!}{\includegraphics[angle=-90]{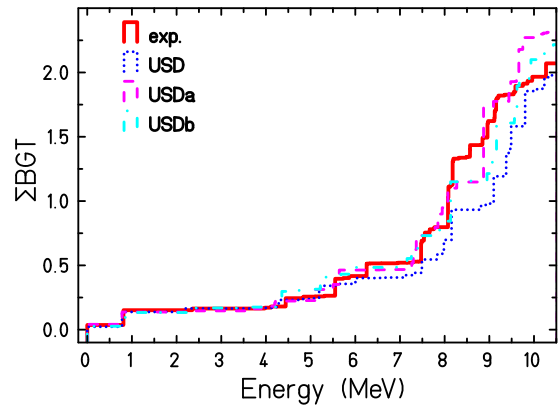}}
\caption[]{(Color online) Summed Gamow-Teller strength distribution for the decay of $^{33}$Ar. The experimental distribution 
           is compared to shell-model calculation by B.A. Brown~\cite{brown09}
           with the universal $sd$ shell interaction USD and two updated versions of the same interaction~\cite{brown06}.
           For these calculations, we used a quenching factor of 0.5, the generally accepted value. The experimental data
           correspond only to resolved proton groups obtained from the
           average values of Table~\ref{tab:bgt} .
	   }
\label{fig:bgt}
\end{center}
\vspace*{-0.9cm}
\end{figure}

Using this quenching factor, we can compare the sum\-med experimental B(GT) strength with the prediction from the shell model. For this comparison, 
we use calculations with three different effective shell-model interactions for the $sd$ model space: the USD interaction~\cite{brown88} and
two newly determined interactions USDa and USDb~\cite{brown06}. For this purpose, we convert the experimental branching ratio into a B(GT) strength for each
individual level observed. As a function of the excitation energy, this strength is then summed. The result is shown in figure~\ref{fig:bgt}.
At low excitation energies, almost perfect agreement of the experimental data with all three interactions is obtained. At higher excitation 
energy, the new interactions USDa and USDb seem to slightly better fit the data. Nonetheless, for all three interactions, the agreement is
remarkable.

\begin{figure}
\begin{center}
\resizebox{.42\textwidth}{!}{\includegraphics{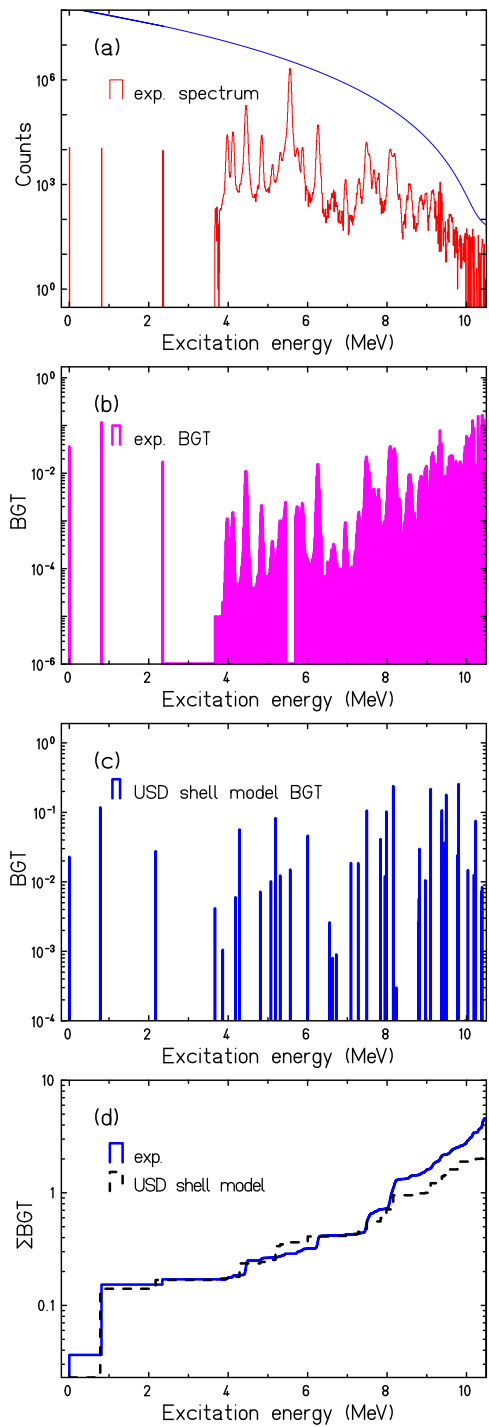}}
\caption[]{(Color online) (a) Proton and $\gamma$-ray spectrum transformed into a spectrum of excitation energies. The three transitions to bound states
           are the discrete lines at low energies. The solid line shows the statistical rate function (arb. units). 
           (b) The spectrum gives the $\gamma$ and proton transitions converted into the Gamow-Teller strength of the individual 
           experimental transitions (see text for details). The contribution of the IAS is removed. (c)  
           Gamow-Teller strength for individual transitions as calculated by means of shell-model calculations with the USD interaction.
           (d) Integrated Gamow-Teller strength as a function of the excitation energy as determined from the
           proton spectrum by adding the three transitions to bound states for the experimental spectrum (full line) and the shell-model
           calculations with the USD interaction (dashed line). 
	   }
\label{fig:bgt_spec}
\end{center}
\end{figure}

The procedure just described has one drawback: Only decay strength identified as a peak usually is taken into account. Very weak decay 
branches are omitted. Therefore, we developed a procedure which uses all strength present in the proton spectra. For this purpose, we 
take the total charged-particle spectrum as obtained by means of the silicon detectors and subtract first the $\beta$ background by a 
fit of the low-energy part of this spectrum. In a second step, we subtract the decay strength to the first and second excited state. 
These contributions are obtained from the proton spectrum conditionned by the detection of these $\gamma$ rays renormalized by the 
gamma efficiency.
This yields the proton decay strength to the ground state. Now we have to convert this spectrum bin by bin to a center-of-mass (CM) 
spectrum. To this we add the decay spectrum (again converted to the center-of-mass) to the first excited state shifted by the excitation 
energy of this state. Similarly, we treat the spectrum for decays to the second excited state. After a correction for the proton 
separation energy, this yields the excitation energy spectrum to which we have added the decay to bound states (see 
figure~\ref{fig:bgt_spec}a). After calculating the statistical rate function $f$ for each energy bin, we can determine the B(GT) value 
for each energy bin (figure~\ref{fig:bgt_spec}b). This spectrum can be compared to the theoretical B(GT) values for individual 
transitions (figure~\ref{fig:bgt_spec}c) and the summed B(GT) distribution (figure~\ref{fig:bgt_spec}d). At high excitation energies, 
the summed experimental B(GT) distribution exceeds the theoretical one with a quenching of 0.5. If we integrate the B(GT) up to 11~MeV
we get 5.3 which correspond to a quenching of 0.59. To discuss the contribution at very high excitation energy is always delicate, 
enough to say that if we integrate in the full Q$_EC$ window of 11.6193~MeV the cumulated B(GT) grows to 41.2.

A possible explanation for this large B(GT) sum is that we shift too much strength to high excitation energies by attributing
too much decay strength to decays to excited states, in particular to the second excited state.
To test this assumption, we calculated the GT strength distribution also by omitting the second excited state and both excited states. 
However, even when omitting both excited states, we still exceed the quenched theoretical B(GT) values.

Another possible explanation could be that we deal to some extent with forbidden transitions. In the latest analysis of Singh et al.~\cite{singh98a},
the log(ft) values of allowed transitions are distributed over a large range from 3 to 7, whereas forbidden transitions start with values
of about 5. As can be seen in figure~\ref{fig:scheme}, we determined indeed high log(ft) values which could be due to forbidden transitions.
However, as we have no means to distinguish between allowed and forbidden transitions, we assume that all transitions are allowed, but we
give their spins in parenthesis to indicate this uncertainty.

Finally, background is of course also a concern. Proton-proton pile up might create high-energy events which get an important weight due 
to the larger space phase factor. However, we do not have any means to distinguish such possible background counts from real counts.

The good agreement between experimental and theoretical B(GT) strength up to rather high excitation energies shows that the quenching
is the same at low and at high excitation energies. A comparison of figures~\ref{fig:bgt} and~\ref{fig:bgt_spec} shows that no major 
proton groups are missing in table~\ref{tab:protons} and basically all strength is identified in the peaks. If this were not the case, 
the second procedure arriving at figure~\ref{fig:bgt_spec} should yield more B(GT) strength than the procedure which produces the 
spectrum of figure~\ref{fig:bgt}.

\subsection{Spins and parities}

Under the assumption of allowed transitions, all levels populated in the $\beta$ decay of $^{33}$Ar (I$^\pi$~= 1/2$^+$) have spin/parity 
1/2$^+$ or 3/2$^+$. However, certain characteristics observed in the present study may allow to distinguish between these two 
possibilities. We will use the observed proton lines to propose spins for some levels. This will be discussed in the following.

Proton emission from a nuclear state is governed by two effects: the barrier penetration and the spatial overlap of the initial and the final states.
For a given nucleus, the barrier penetration depends on the energy available for the proton and the barrier height. The contribution of the Coulomb
barrier is the same for all states and all emissions. However, for a transition with a non-zero angular momentum, the angular momentum
barrier is added. Therefore, proton transitions to the ground state are favored by the available energy, except if angular momentum has
to be "carried away" by the proton. If states emit protons only to an excited state in the proton daughter nucleus, this can be an indication
that this transition has a lower angular momentum than the ground state transition, a fact which would enable us to distinguish between two 
possible spin assignments.

Such an assignment, however, neglects the possibility that strongly different overlaps of the initial and final wave functions can have 
a similar effect. In principle, model calculations would have to be performed for all states involved before such an assignment can 
be made. We believe that this is not very meaningful, as the states calculated e.g. in shell-model calculations might differ strongly 
from the observed states, even if the excitation energy is equivalent.

We assign tentatively a spin of 3/2$^+$ to all states which decay only to the first excited state in the proton-daughter nucleus
$^{32}$S. We give these spin/parities in parenthesis to make clear that this is a tentative assignment.

The states at 7762~keV and at 8491~keV decay to the 0$^+$ ground state and the 0$^+$ second excited state. This indicates that the 
emitting state has most likely a spin/parity of 1/2$^+$ which yields an $\ell$~= 0 proton emission.

For other states, barrier penetration calculations (see table~\ref{tab:penetration}) may elucidate the situation. The barrier penetration 
is calculated from Coulomb wave functions~\cite{brown91} which yields partial half-lives for both spin possibilities for decays to
the ground and excited states. The ratios of the partial half-lives for emission to the ground and excited states for both spins are 
then compared to the ratio of the experimental branching ratios. If we believe that the situation is relatively clear and one 
spin assignment gives a much closer agreement than the other spin, we propose a spin. The procedure used is tested with and works well 
for the state at 7292~keV, for which we find a spin of 3/2$^+$ in agreement with the literature~\cite{endt98}.

\renewcommand{\baselinestretch}{0.5}
\begin{table*}[hht]
\begin{center}
\mbox{\footnotesize
\begin{tabular}{ccccccccc}
\hline
\hline
 E$_x$  &&  E$_p$ &  angular  &   possible spin & T$_{1/2}$ && BR   & proposed  \\
 (keV)  &&  (keV) &  momentum &     of level    &     (s)   && (\%) &  spin \\
\hline
5310 &&	2941 &	l=0 & 1/2 &  0.27e-21 & &  0.0769  &  3/2   \\
	 &&	 762 &	l=2 &     &  0.58e-16 & &  0.0201  & \\
	 &&	2941 &	l=2 & 3/2 &  0.50e-20 & &          & \\
	 &&	 762 &	l=0 &     &	 0.74e-18 & &          & \\
&&&&&&&\\		
6253 &&	3857 & 	l=0 & 1/2 &  0.13e-21 & &  0.722   & \\
	 &&	1691 &	l=2 &     &  0.12e-18 & &  0.0328  & \\
	 &&	3857 &	l=2 & 3/2 &  0.14e-20 & &          & \\
	 &&	1691 &	l=0 &     &  0.29e-20 & &          & \\
&&&&&&&\\		
7292 &&	4860 &	l=0 & 1/2 &  0.76e-22 & &  0.0108  &  3/2    \\
	 &&	2710 &	l=2 &     &  0.75e-20 & &  0.00691 & \\
	 &&	4860 &	l=2 & 3/2 &  0.55e-21 & &          & \\
	 &&	2710 &	l=0 &     &  0.36e-21 & &          & \\
&&&&&&&\\
7475 &&	5039 &	l=0 & 1/2 &  0.71e-22 & &  0.244   & \\
	 &&	2886 &	l=2 &     &  0.55e-20 & &  0.0392  & \\
	 &&	5039 &	l=2 & 3/2 &  0.48e-21 & &          & \\
	 &&	2886 &	l=0 &     &  0.29e-21 & &          & \\
&&&&&&&\\		
7666 &&	5225 &	l=0 & 1/2 &  0.67e-22 & &  0.0297  & \\
	 &&	3066 &	l=2 &     &  0.40e-20 & &  0.00438 & \\
	 &&	5225 &	l=2 & 3/2 &  0.42e-21 & &          & \\
	 &&	3066 &	l=0 &     &  0.24e-21 & &          & \\
&&&&&&&\\		
8077 &&	 5623 &  l=0 & 1/2 &  0.59e-22 &&   0.154  & \\
	 &&  3469 &  l=2 &     &  0.22e-20 &&   0.0530 & \\
	 &&  5623 &  l=2 & 3/2 &  0.32e-21 &&          & \\
	 &&  3469 &  l=0 &     &  0.16e-21 &&          & \\
&&&&&&&\\	
8179 &&	5723 &  l=0 & 1/2 &  0.57e-22 & &  0.103   &  1/2   \\
	 &&  3576 &  l=2 &     &  0.20e-20 &&   0.0085  & \\
	 &&  5723 &  l=2 & 3/2 &  0.30e-21 &&           & \\
	 &&  3576 &  l=0 &     &  0.15e-21 &&           & \\
&&&&&&&\\
8557 &&  6100 &  l=0 & 1/2 &  0.52e-22 & &  0.0146  &  3/2   \\
	 &&  3926 &  l=2 &     &  0.13e-20 & &  0.00812 & \\
	 &&  6100 &  l=2 & 3/2 &  0.25e-21 & &          & \\
	 &&  3926 &  l=0 &     &  0.12e-21 & &          & \\
&&&&&&&\\	
8967 &&	6480 &  l=0 & 1/2 &  0.48e-22 &  & 0.0103   &  1/2    \\
	 &&  4330 &  l=2 &     &  0.86e-21 & &  0.00142  & \\
	 &&  6480 &  l=2 & 3/2 &  0.20e-21 & &           & \\
	 &&  4330 &  l=0 &     &  0.96e-22 & &           & \\
&&&&&&&\\		
9119 &&	6628 &  l=0 & 1/2 &  0.46e-22 &  & 0.00170  &  3/2   \\
	 &&  4474 &  l=2 &     &  0.75e-21 & &  0.00360 & \\
	 &&  6628 &  l=2 & 3/2 &  0.19e-21 & &          & \\
	 &&  4474 &  l=0 &     &  0.90e-22 & &          & \\
&&&&&&&\\	
\hline
&&&&&&&\\	   
9152 &&	6657 &  l=0 & 1/2 &  0.46e-22 &  & 0.00050   & 3/2   \\
	 &&  4505 &  l=2 &     &  0.73e-21 & &  0.00466 & \\
	 &&  3016 &  l=0 &     &  0.25e-21 & &  0.00066 & \\
&&&&&&&\\    
	 &&  6657 &  l=2 & 3/2 &  0.19e-21 & &           & \\
	 &&  4505 &  l=0 &     &  0.88e-22 & &           & \\
	 &&  3016 &  l=2 &     &  0.44e-20 & &          & \\
\hline
\hline
\end{tabular}
}
\end{center}
\renewcommand{\baselinestretch}{1.0}
\vspace*{-0.5cm}
\caption{Barrier penetration calculations used to tentatively assign spins to certain levels. The first column gives 
          the excitation energy of the state, followed by the two or three proton energies from this state. The next 
          two columns give the angular momentum used in the calculations which is deduce from the two spin possibilities of the proton
          emitting state, 1/2 or 3/2, given in the following column. The next column is the barrier penetration half-life calculated
          with the angular momentum l=0 or l=2. 
          The sixth column is the experimental branching ratio of the two/three proton transitions from the 
          state. These branching ratios are given only once for each proton group. For some states, in the 
          final column, we give our tentative spin assignments.}
\label{tab:penetration}
\end{table*}

For the state at 9152~keV which decays to the three lowest states of $^{32}$S, we determine that for a 3/2$^+$ emitting state the decay to the 
first excited state is indeed the fastest transition thus receiving the highest branching ratio. For a spin 1/2$^+$, the decay to the ground state
is calculated to be a factor of ten faster than the decay to the other two states, opposite to the experimental observation. We therefore attribute
a spin 3/2$^+$ to this state.

We underline again that the present spin assignments have to be confirmed by other means before they should be accepted. Therefore, 
they appear in parenthesis in figure~\ref{fig:scheme}.

\subsection{Search for candidate states for isospin mixing}

Proton emission from the T=3/2 IAS in $^{33}$Cl can only take place due to isospin impurity either in the proton emitting state or 
in the proton daughter state. Therefore, Honkanen et al.~\cite{honkanen96}  searched for states in the vicinity of the IAS with 
which the IAS can mix. We identified three of the four states discussed by Honkanen et al. There is one state about 230~keV 
below the IAS and two other states 180~keV and 310~keV above the IAS. However, we do not have evidence for a state about 
100~keV below the IAS which would be the best candidate for mixing between such a state (T=1/2, I$^\pi$=1/2$^+$) and 
the IAS (T=3/2, I$^\pi$=1/2$^+$). 

The proton group identified by Honkanen et al.~\cite{honkanen96} at 3073~keV corresponds to our peak at 3066~keV. However, 
we clearly identify the proton group to decay to the first excited state and not to the ground state as suggested by Honkanen 
et al. Therefore, the emitting state is no longer in the vicinity of the IAS.

We do not completely exclude the possibility of such a close-by state, as our resolution is maybe not good enough to resolve 
several small peaks in the close vicinity of the strong IAS. However, if such a state is absent, the mixing has to take place 
with states further away from the IAS.  A better resolution experiment with higher statistics is certainly needed to answer 
this question.

A state with (T=1/2, I$^\pi$=1/2$^+$) about 100~keV below the IAS was also observed in proton scattering experiments on 
$^{32}$S~\cite{aleonard76}. This would of course be a good candidate for isospin mixing with the IAS. As mentioned above, 
we do not observe such a state. However, we cannot exclude that our proton peak at 3066~keV contains also a small branch to 
the ground state of $^{32}$S.

It is interesting to note that the shell model using the three different USD interactions~\cite{brown85,brown06} does not predict
any (T=1/2, I$^\pi$=1/2$^+$) state close to the IAS. The states close to the IAS have all a spin of I$^\pi$=3/2$^+$.
The closest I$^\pi$=1/2$^+$ state is at about 400~keV above the IAS for the USD interaction, 550~keV above for the USDa interaction, 
and 450~keV above the IAS for USDb. Below the IAS, the first (T=1/2, I$^\pi$=1/2$^+$) state is typically 800-900 keV away 
from the IAS.

\subsection{Decay scheme of $^{33}$Ar}

\begin{figure*}[tth]
\begin{center}
\vspace*{-0.4cm}
\resizebox{.75\textwidth}{!}{\includegraphics{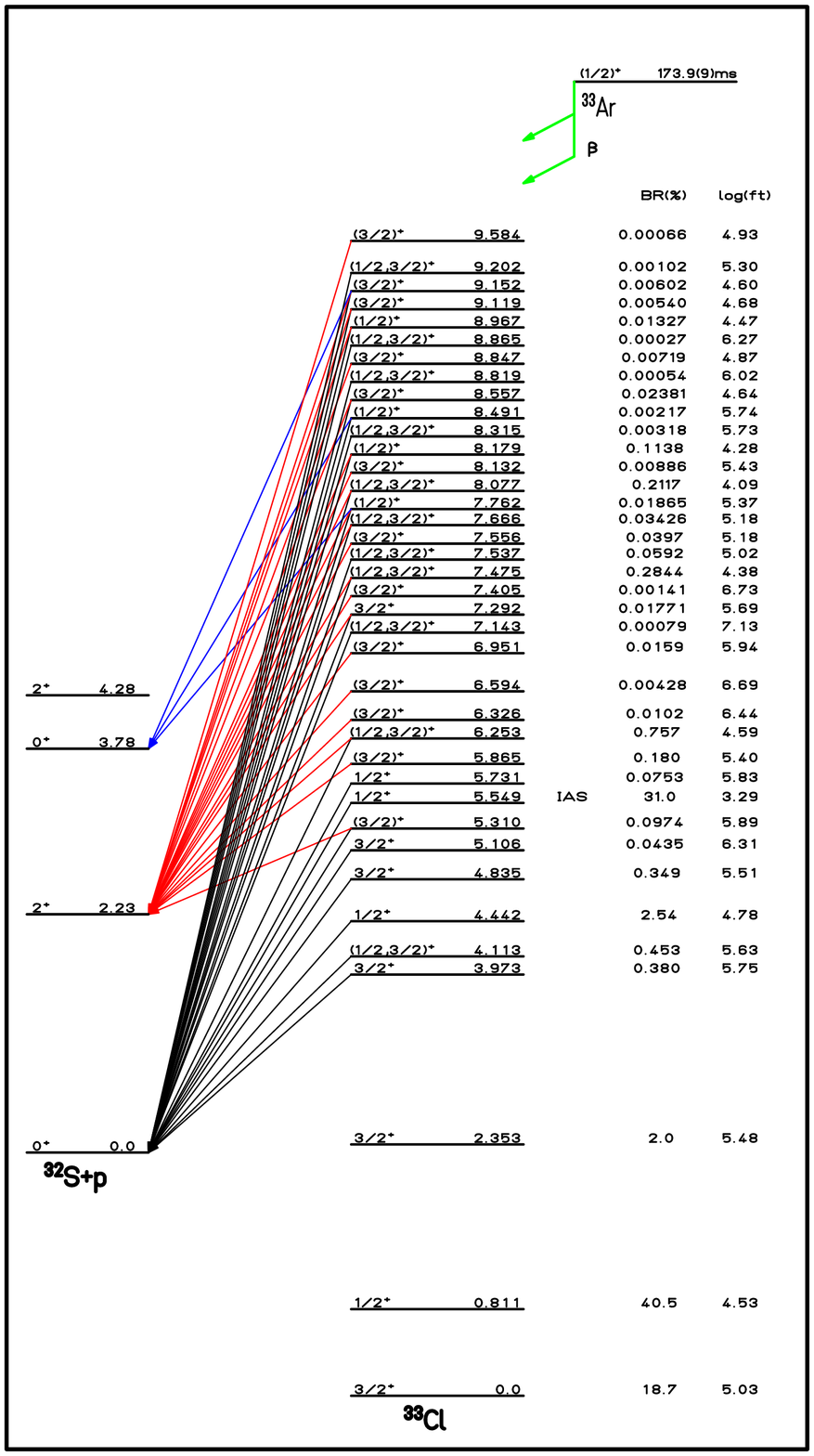}}
\caption[]{(Color online) Decay scheme of $^{33}$Ar. All the decay branches, either to particle-bound states or to proton-unbound states decaying by 
           proton emission to the three lowest states of $^{32}$S, are shown. The excitation energy of the decaying state, the 
           branching ratio and the log(ft) value are given. Spins tentatively assigned in the present work to be 1/2$^+$ or 3/2$^+$ 
           are given in parentheses. The spins of levels for which both spins (1/2$^+$, 3/2$^+$) are given in parenthesis are 
           determined from $\beta$ decay selection rules assuming allowed transitions. Other spin/parities are from Ref.~\cite{endt98}.
	   }
\label{fig:scheme}
\end{center}
\end{figure*}

The information accumulated in the present paper together with the results from previous work enables us to establish a rather complete 
decay scheme of $^{33}$Ar with decays to bound states and proton unbound states which then decay to the three lowest states 
of $^{32}$S by proton emission. Figure~\ref{fig:scheme} gives this decay scheme. We remind the reader that the branching ratios for the three
proton-bound states are based on the assumption of mirror symmetry of the ground states of $^{33}$Ar and $^{33}$P.

\section{Summary}

Using a novel experimental setup of high granularity and high efficiency for charged-particle detection combined with 
a high-efficiency $\gamma$ array allowed a complete study of the decay of $^{33}$Ar to be performed. The experimental 
results are to a large extent in agreement with previous experimental studies of this nucleus. They enabled us to 
establish a quasi complete decay scheme of $^{33}$Ar and to compare the experimental results with shell-model 
calculations using different effective interactions optimized for the $sd$ shell-model space. The excellent agreement obtained 
testifies of the high quality of the shell model in this region of the chart of nuclei. The comparison of experimental B(GT) 
strength to well identified levels and theoretical calculations yields a quenching factor of 0.49(4), in excellent agreement 
with the accepted value. States in close vicinity of the isobaric analog state of the $^{33}$Ar ground state in $^{33}$Cl are 
searched for and discussed in terms of isospin mixing. However, a state identified before could not be observed in the present work.

\section*{Acknowledgment}

We would like to thank the whole GANIL and, in particular, the accelerator staff for their support during 
the experiment. We express our gratitude to the EXOGAM collaboration for providing us with the germanium 
detectors. This work was partly funded by the Conseil r\'egional d'Aquitaine and the EU 
through the Human Capital and Mobility program. 
We acknowledge support from CICYT via contract FPA2007-62179. RDR was supported via a FPI grant.
We are in debt of B.A. Brown for providing us with the shell-model calculations.

\end{document}